\documentclass[useAMS]{mn2e}
\usepackage{graphicx}                   
\input{epsf.sty}                        
\input{psfig.sty}

\title[1E 1207.4-5209: a strange star?]
{1E 1207.4-5209: a low-mass bare strange star?}

\author[Xu]{R.X.~Xu \\
School of Physics, Peking University, Beijing 100871, China;
Email: {\tt rxxu@bac.pku.edu.cn}}

\date{Received~~~~~; Accepted}

\pagerange{\pageref{firstpage}--\pageref{lastpage}}
\pubyear{2004}

\begin{document}

\maketitle

\label{firstpage}

\begin{abstract}

Both rotation- and accretion-powered low-mass bare strange stars
are studied, the astrophysical appearances of which are especially
focused. It is suggested that low-mass bare strange stars, with
weaker ferromagnetic fields than that of normal pulsars, could
result from accretion-induced collapses (AIC) of white dwarfs.
According to its peculiar timing behavior, we propose that the
radio-quiet object, 1E 1207.4-5209, could be a low-mass bare
strange star with polar surface magnetic field $\sim 6\times
10^{10}$ G and a few kilometers in radius.
The low-mass bare strange star idea is helpful to distinguish
neutron and strange stars, and is testable by imaging pulsar-like
stars with the future Constellation-X telescope.

\end{abstract}


\section{Introduction}

Astrophysics offers an alternative channel for us to explore the
fundamental laws in the nature, and the study of quark stars obeys
exactly this spirit. It is a clear goal for part of laboratory
physicists to find quark-gluon-plasma (or quark matter) in order
to research into the problem of the elemental color interaction,
whereas to detect astrophysical quark matter may be a shortcut.
Though one may conventionally think that pulsars are ``normal''
neutron stars \cite{lp04}, it is still an open issue whether
pulsar-like stars are neutron or quark stars
\cite{glen00,weber99,mad99}, since {\em no} convincing work,
neither in theory from first principles nor in observation, has
confirmed Baade-Zwicky's original idea that supernovae produce
neutron stars.
Therefore, the question of detecting astrophysical quark matter is
changed to be: how to identify a quark star?

One kind of frequently discussed quark stars are those with
strangeness, namely strange stars, which are very likely to exist.
There are a few ways (e.g., of cooling behaviors, mass-radius
relations, etc.) proposed, by which neutron and strange stars
could be distinguished.
However the peculiar nature of quark surface has not been noted
until 1998 \cite{usov02,xu03c}.
As the strange star model \cite{xu03d} may work well to understand
the various observations of pulsar-like stars, including glitches
and free-precessions \cite{z04}, there are at least three natural
motivations to study such stars with low-masses.

(1) The formation of low-mass strange stars is a direct
consequence of the presumption that pulsar-like stars are actually
quark stars rather than neutron stars. One can {\em not} rule out
this possibility (but neglected unfortunately) now from either
first principles or astrophysical observations. This paper is
trying to catch astrophysicist's attention to the investigation
relevant.

(2) ``Low-mass'' may helpful for identifying strange stars.
Bare strange stars can be very low-massive with small radii, while
normal neutron stars can not.
It is well known that the masses and radii of neutron and strange
stars with almost the maximum mass are similar\footnote{
This is the reason that one generally believes that neutron and
strange stars can not be distinguished by measuring only their
masses {\em or} radii, but tries to compare the
observation-determined mass-radius relations with the theoretical
ones in order to identify strange stars.
}; %
nonetheless, low-mass neutron and strange stars have remarkably
different radii \cite{afo86,bom97,lxd99}.
Due to the color confinement by itself rather than gravitational
bounding, a bare strange star could be very small, the radius of
which with mass $M\la M_\odot$ is, from Eq.(\ref{mr}),
\begin{equation}
R = 1.04\times 10^6{\bar B}_{60}^{-1/3}(M/M_\odot)^{1/3}~~{\rm cm},%
\label{R}
\end{equation}
and $R=(4.8, 2.3, 1.0)$ km for $M=(10^{-1}, 10^{-2},
10^{-3})M_\odot$ if the bag constant ${\bar B}=60$ MeV/fm$^3$. But
the radii of neutron stars with $\sim 0.5M_\odot$ are generally
greater than 10 km, and the minimum mass of a stable neutron star
is $\sim 10^{-1}M_\odot$, with radius $R\sim 160$ km (about two
orders larger than that of low-mass bare strange stars with
similar masses) \cite{st83}.
It is consequently possible that we can distinguish neutron and
strange stars by direct measurements of the radii\footnote{%
Also the very distinguishable mass-radius ($M-R$) relations of
neutron and quark stars are helpful. The study of Lane-Emden
equation with $n=2/3$ (corresponding to the state of
non-relativistic neutron gas with low masses) results in $M\propto
R^{-3}$, whereas, for low-mass quark stars, $M\propto R^3$, due to
the color confinement of quark matter.
} %
of low-mass pulsar-like stars by X-ray satellites.
Fortunately, low-mass neutron stars have been cared recently
\cite{chp03}, the radii of which may correlate with that of
$^{208}$Pb because of the stellar central densities to be near the
nuclear-matter saturation density.
It is then worth studying low-mass strange stars in order to
obtain crucial evidence for quark stars.

(3) The conventional method to estimate the polar magnetic fields
of radio pulsars has to be modified if no certain reason forces us
to rule out the possibility that low-mass bare strange stars could
exist in the universe (see \S3.1 for details).

Additionally, the identification of a low-mass strange star, with
mass $\la 0.1M_\odot$, may also tell us whether the star is {\em
bare}, since the radius of a bare strange star is much smaller
than that of a crusted one in the low-mass limit
\cite{bpv04,xu03a}.

The layout of the rest of this paper is as follows. First, a
phenomenological view of strange quark matter is introduced in
\S2. After a study of the general nature of rotation-powered as
well as accretion-driven low-mass bare strange stars in \S3, we
focus our attention on the central compact object 1E 1207.4-5209
in \S4, with some investigations of other potential candidates in
\S5. Although the major points of the paper are to propose
candidates of low-mass quark stars according to astrophysical
observations, an effort of probing into the origin of such stars
is tried in \S6. Finally, conclusions and discussions are
presented in \S7.

\section{Quark matter phenomenology}

Strange quark stars are composed of quark matter with almost equal
numbers of $u$, $d$, and $s$ quarks.
There are actually two {\em different} kinds of quark matter to be
investigated in lab-physics and in astrophysics, which appear in
two regions in the quantum chromo-dynamics (QCD) phase diagram
(Fig.1).
Quark matter in lab-physics and in the early universe is
temperature-dominated (temperature $T\gg 0$, baryon chemical
potential $\mu_{\rm B}\sim 0$), while that in quark stars or as
cosmic rays is density dominated ($T\sim 0$, $\mu_{\rm B}\gg 0$).
Previously, Monte Carlo simulations of lattice QCD (LQCD) were
only applicable for cases with $\mu_{\rm B}=0$. Only recent
attempts are tried at $\mu_{\rm B}\neq 0$ (quark stars or nuggets)
in LQCD. We have then to rely on phenomenological models to
speculate on the properties of density-dominated quark matter.

In different locations of the diagram (Fig.1), besides that the
interaction strength between quarks and gluons is weak or strong,
the vacuum would have different features and is thus classified
into two types: the perturbative-QCD (pQCD) vacuum and
nonperturbative-QCD (QCD) vacuum. The coupling is weak in the
former, but is strong in the later.
Quark-antiquark (and gluons) condensations occur in QCD vacuum
(i.e., the expected value of $\langle {\bar q}q\rangle \neq 0$),
but not in pQCD vacuum.
The chiral symmetry is spontaneously broken in case the vacuum is
changed from pQCD to QCD vacuums, and quarks become then massive
constituent ones.
LQCD calculations \cite{kog91} show that the value of $\langle
{\bar q}q\rangle$ increases when the color coupling becomes strong
(i.e., as the temperature or the baryon density decrease).
Therefore, we note that the quark de-confinement and the chiral
symmetry restoration may {\em not} take place at a same time.
\begin{figure}
  \centering
  \begin{minipage}[t]{.5\textwidth}
    \centering
    \includegraphics[width=3in]{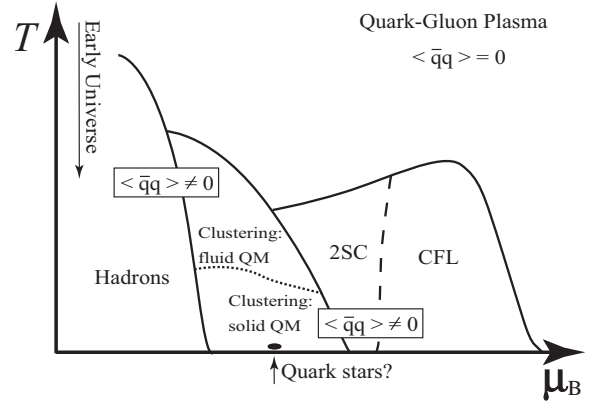}
  \end{minipage}
     \caption{
Schematic illustration of QCD phase diagram.}%
\end{figure}

Considerable theoretical efforts have been made in past years to
explore the QCD phase diagram.
When $T$ or $\mu_{\rm B}$ are extremely high, there should be
quark-gluon plasma (QGP) phase because of the asymptotic freedom,
and vacuum is of pQCD.
However, in a relatively lower energy limit, especially in the
density-dominated region, the vacuum is phase-converted to QCD one
but the quarks could be still deconfined. It is a hot point to
investigate the possibility that real quarks may also be condensed
(i.e., $\langle q q\rangle \neq 0$) simultaneously when $\langle
{\bar q}q\rangle \neq 0$, the so-called color-superconducting
(CSC) phases (for recent reviews, see, e.g., Ren~\cite{ren04},
Rischke~\cite{ris04}). Actually two CSC phases are currently
discussed. One corresponds to Cooper pairing among the two flavors
of quarks ($u$ and $d$) only, the two-flavor color
superconductivity (2SC) phase, in case that $s$ quark is too
massive to participate. Another one occurs at higher $\mu_{\rm B}$
in which $s$ quarks are relatively less massive and are thus
involved in Cooper pairing, the color-flavor locked (CFL) phase.

However, another possibility can not be ruled out:  $\langle q
q\rangle = 0$ while $\langle {\bar q}q\rangle \neq 0$.
When $T$ is not high, along the reverse direction of the $\mu_{\rm
B}$ axis, the value of $\langle {\bar q}q\rangle$ increases, and
the color coupling between quarks and gluons becomes stronger and
stronger.
The much strong coupling may favor the formation of $n-$quark
clusters ($n$: the number of quarks in a cluster) in the case
\cite{xu03b}. Such quark clusters could be very likely in an
analogy of $\alpha$ clusters moving in nuclei, which are well
known in nuclear physics.
Recent experimental evidence for multi-quark ($n > 3$) hadrons may
increase the possibility of quark clustering.
The clusters are localized\footnote{
We apply ``local'' to refer that ``quark wavefunctions do almost
not overlap''. In this sense, localized clusters can still move
from place to place when $T$ is high, but could be solidified at
low $T$.
} %
to become {\em classical} (rather than quantum) particles when the
thermal de Broglie wavelength of clusters $\lambda\sim
h/\sqrt{3mkT}<l\sim [3n/(4\pi fn_{\rm b})]^{1/3}$ ($m$: the mass
of clusters, $l$: the mean cluster distance, $n_{\rm b}$: the
baryon number density, $f$: quark flavor number), assuming no
interaction is between the clusters.
Calculation based on this inequality for $f=3$ shows that cluster
localization still exists even in temperature $T\sim 1$ MeV if
$n\sim 10^2$. In addition, the interaction in-between, which is
neglected in the inequality, would also favor this localization.

In case of negligible interaction, quark clusters would become a
quantum system in case of low temperature. However, the
interaction is certainly not weak since the vacuum is of QCD
($\langle {\bar q} q \rangle\neq 0$).
Now, a {\em competition} between condensation and solidification
appears then, just like the case of laboratory low-temperature
physics.
Quark matter would be solidified as long as the interaction energy
between neighboring clusters is much larger than that of the
kinetic thermal energy.
This is why {\em only helium}, of all the elements, shows
superfluid phenomenon though other noble elements have similar
weak strength of interaction due to filled crusts of electrons.
The essential reason for the occurrence of CSC is that there is an
attractive interaction between two quarks at the Fermi surface.
But, as discussed, much strong interaction may result in the quark
clustering and thus a solid state of quark matter.
In conclusion, a new phase with $\langle {\bar q} q\rangle\neq 0$
but $\langle qq \rangle=0$ is suggested to be inserted in the QCD
phase diagram (Fig.1), which could exist in quark stars.

Astrophysics may teach us about the nature of density-dominated
quark matter in case of those theoretical uncertainties.
There are at lest two astrophysical implications of the solid
quark matter state proposed.

(1) Quark stars in a solid state can be used to explain naturally
the observational discrepancy between glitches and
free-precessions of radio pulsars \cite{link03}, since a solid
quark star is just a rigid-like body (no damping precession), and
glitches would be the results of star-quakes. Actually, it is
found \cite{z04} that the general glitch natures (i.e., the glitch
amplitudes and the time intervals) could be reproduced if the
solid quark matter has properties of shear modulus $\mu=10^{30\sim
34}$ erg/cm$^3$ and critical stress $\sigma_{\rm c}=10^{18\sim
24}$ erg/cm$^3$.

(2) Ferro-magnetization may occur in a solid quark matter, without
field decay in the stars.
Magnetic field plays a key role in pulsar life, but there is still
no consensus on its physical origin although some ideas relevant
(e.g., the flux conservation during collapse, the dynamo action)
appeared in the literatures.
Whereas, an alternative suggestion is the generation of strong
magnetic fields by spontaneously broken ferromagnetism in quark
matter \cite{tat00}. One of the advantages of ferromagnetic origin
could be their unchangeable nature since there is no convincing
evidence that fields decay in isolate pulsar-like stars\footnote{
The seeming field-decay in the $P-{\dot P}$ diagram could be
arisen from known selection effects, based on the simulations
\cite{itoh92}.
}. %
However, the magnetic domain structure may be destroyed by the
turbulent motion in a fluid quark star. This worry does not exist
if pulsars are solid quark stars \cite{xu03b}.
Quark clusters with magnetic momentum may exist in solid quark
stars. Solid magnetic quark matter might then magnetize itself
spontaneously at sufficient low temperature (below its Curie
critical temperature) by, e.g., the flux-conserved field.
Ferromagnetism saturated may result in a very strong {\em dipole}
magnetic field.
We therefore presume a ferromagnetic origin of pulsar fields in
the following calculations.

\section{Low-mass bare strange stars}

\subsection{Rotation-powered phase}

The energy conservation for an orthogonal star (i.e., the
inclination angle between magnetic and rotational axes is
$\alpha=90^{\rm o}$) with a magnetic dipole moment $\mu$, a moment
of inertia $I$, and angular velocity $\Omega$ gives
\begin{equation}
{\dot \Omega} = -{2\over 3Ic^3}\mu^2\Omega^3.%
\label{dotO}
\end{equation}
This rule keeps quantitatively for any $\alpha$, as long as the
braking torques due to magnetodipole radiation and the unipolar
generator are combined \cite{xq01}.
For a star with polar magnetic field $B$ and radius $R$, the
magnetic moment
\begin{equation}
\mu={1\over 2}BR^3,%
\label{B}
\end{equation}
if the fields are in pure dipole magnetic configuration or if the
star is an uniformly magnetized sphere. This results in the
conventional magnetic field derived from $P$ and $\dot P$
($P=2\pi/\Omega$ is the spin period),
\begin{equation}
B = 6.4\times 10^{19}\sqrt{P{\dot P}}~~{\rm G}, %
\label{b}
\end{equation}
if ``{\em typical}'' values $I=10^{45}$ g$\cdot$cm$^2$ and
$R=10^6$ cm are assumed. Note: the field is only half the value in
Eq.(\ref{b}) if one simply suggests $\mu=BR^3$ \cite{mt77}.

\begin{table*}
\begin{center}
\begin{tabular}{lllcclll}
\hline Pulsars & $P$ (ms) & $\dot P$ (s/s) & $M$ ($M_\odot$) &
$\mu_{\rm m}$ (model 1) & $\mu_{\rm m}$ (model 2) & $\mu_{\rm v}$
(model 3) & $\mu_{\rm v}$ (model 4) \\ \hline
J1518$^{\rm a}$ & 40.94 & 2.73E-20  & 1.56  & 4.27E-7 & ~~~~--- & 2.31E+8 & ~~~~--- \\
B1534$^{\rm a}$ & 37.90 & 2.42E-18 & 1.34  & 4.00E-6 & 3.10E-6 & 2.10E+9 & 3.52E+9 \\
B1913$^{\rm a}$ & 59.03 & 8.63E-18 & 1.44  & 9.25E-6 & 6.89E-6 & 4.98E+9 & 8.99E+9 \\
B2127$^{\rm a}$ & 30.53 & 4.99E-18 & 1.35  & 5.15E-6 & 3.97E-6 & 2.72E+9 & 4.57E+9 \\
B2303$^{\rm a}$ & 1066 & 5.69E-16 & 1.30  & 3.28E-4 & 2.56E-4 & 1.71E+11 & 2.80E+11 \\
J0737A$^{\rm b}$ & 22.70 & 1.74E-18 & 1.34  & 2.62E-6 & 2.03E-6 & 1.38E+9 & 2.31E+9 \\
J0737B$^{\rm b}$ & 2773 & 0.88E-15 & 1.25  & 6.65E-4 & 5.19E-4 &
3.43E+11 & 5.61E+11 \\ \hline
\end{tabular}
\end{center}

\caption{Pair neutron stars and the model parameters derived.
Models numbered 1, 2, 3, and 4 are for $\{a=\mu_{\rm m}, \alpha=1;
b=0, \beta=0; \bar{B}=60$MeV/fm$^3$\}, $\{a=\mu_{\rm m}, \alpha=1;
b=0, \beta=0; \bar{B}=110$MeV/fm$^3$\}, \{$a=0, \alpha=0;
b=4\pi\mu_{\rm v}/3, \beta=3; \bar{B}=60$MeV/fm$^3$\}, and \{$a=0,
\alpha=0; b=4\pi\mu_{\rm v}/3, \beta=3; \bar{B}=110$MeV/fm$^3$\},
respectively, where $\bar{B}$ is the bag constant. The former two
are for constant magnetic momentum $\mu_{\rm m}$ (G$\cdot {\rm
cm}^3\cdot$g$^{-1}$) per unit mass, while the later two for
constant momentum $\mu_{\rm v}$ (G) per unit volume. Note: ``a''
denotes Thorsett \& Chakrabarty (1999), ``b'' denotes Lyne et al.
(2004).}

\label{tab1}

\end{table*}

However, neutron star's $I$ and $R$ change significantly for
different equations of state, or for different masses even for a
certain equation of state \cite{lp01}. This means that the
``typical'' values may actually be not approximately constants.
The inconsistency becomes more serious if pulsar-like stars are in
fact strange quark stars since such a quark star could be as small
as a few hundreds of baryons (strangelets).

Let's compute $P$ and $\dot P$ for a quark star with certain mass
$M$ and radius $R$.
First we approximate the momentum of inertia to be $I\simeq
2MR^2/5$ (i.e., a star with uniform density). This approximation
is allowed for low-mass strange stars \cite{afo86}. In this case,
the magnetic field derived from $P$ and $\dot P$ is then
\cite{xxw01}, from Eq.(\ref{dotO}) and Eq.(\ref{B}),
\begin{equation}
\begin{array}{lll}
B & = & \sqrt{0.6P{\dot P}}c^{3/2}M^{1/2}R^{-2}/\pi\\
& = & 5.7\times 10^{19}M_1^{1/2}R_6^{-2}\sqrt{P{\dot P}}~~{\rm G}, %
\end{array}
\label{b-lowmass}
\end{equation}
where $M_1=M/M_\odot$ and $R=R_6\times 10^6$ cm.
As the strong magnetic fields of pulsars are suggested to be
ferromagnetism-originated, we then assume the magnetic momentum
\begin{equation}
\mu=a M^\alpha+b R^\beta, %
\label{mu}
\end{equation}
where $\{a, \alpha; b, \beta\}$ is a parametric set.
If the magnetized momentum per unit volume is a constant $\mu_{\rm
v}$, one has $\{a=0, \alpha=0; b=4\pi\mu_{\rm v}/3, \beta=3\}$. In
case the magnetized momentum per unit mass is a constant $\mu_{\rm
m}$, one has then $\{a=\mu_{\rm m}, \alpha=1; b=0, \beta=0\}$.
From Eq.(\ref{dotO}) and Eq.(\ref{mu}), one comes to
\begin{equation}
P{\dot P}={20\pi^2(a M^\alpha+b R^\beta)^2\over 3c^3MR^2}.%
\label{ppdot}
\end{equation}
This equation shows that a pulsar with certain initial parameters
($M$ and $R$) evolves along constant $(P{\dot P})$. Integrating
Eq.(\ref{ppdot}) above, one obtains the pulsar age $T$
\begin{equation}
T={P^2-P_0^2\over 2P{\dot P}},%
\label{T}
\end{equation}
where $P_0$ is the initial spin period. The age $T=T_{\rm c}\equiv
P/(2{\dot P})$ when $P_0\ll P$.

The mass of a pulsar is detectable dynamically if it is in a
binary system, but precise mass estimates are only allowed by the
measurement of relativistic orbital effects.
It is therefore to determine the model parameters ($\mu_{\rm m}$
and $\mu_{\rm v}$) in Eq.(\ref{ppdot}) by pair neutron star
systems \cite{tc99,lyne04}.
Four cases possible are investigated, the calculational results of
which are listed in Table 1. The mass-radius relations for the
calculation are in the regime of strange quark matter described by
a simplified version of the MIT bag model, in which the relation
between pressure $\cal P$ and the density $\rho$ is given by
\begin{equation}
{\cal P}={1\over 3}(\rho-4{\bar B}),%
\label{eos}
\end{equation}
where the bag constant $\bar B$ is chosen to be 60 MeV/fm$^3$ (low
limit) and 110 MeV/fm$^3$ (up limit), respectively. It is assumed
that the mass and radius of a star do not change significantly
during quark-clustering and solidification as the star cools.

\begin{figure}
\centerline{\psfig{file=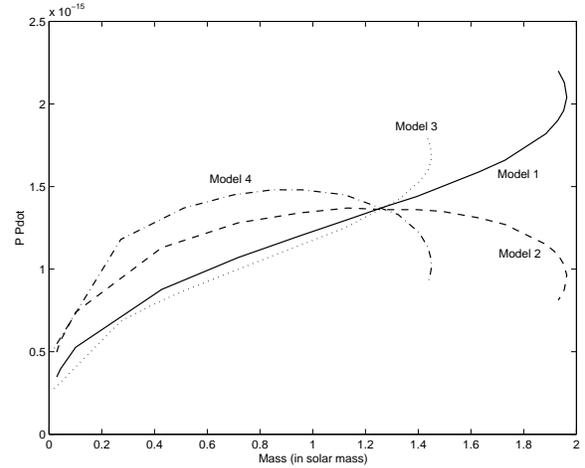,width=3in}}
\caption{The $P{\dot P}$ value based on Eq.(\ref{ppdot}). Model 1:
$\mu_{\rm m}=4.97\times 10^{-4}$ G$\cdot$cm$^3\cdot$g$^{-1}$,
Model 2: $\mu_{\rm m}=3.88\times 10^{-4}$
G$\cdot$cm$^3\cdot$g$^{-1}$, Model 3: $\mu_{\rm v}=2.57\times
10^{11}$ G, and Model 4: $\mu_{\rm v}=4.21\times
10^{11}$ G. %
\label{fig1}}
\end{figure}
The values of $\mu_{\rm m}$ and $\mu_{\rm v}$ are grouped into two
classes in Table 1. One class has higher $\mu_{\rm m}$ or
$\mu_{\rm v}$ for normal pulsars (B2303 \& J0737B only), but the
other (of millisecond pulsars) has lower $\mu_{\rm m}$ or
$\mu_{\rm v}$.
We just average the values of B2303 \& J0737B for indications of
normal pulsars; Model 1: $\mu_{\rm m}=4.97\times 10^{-4}$
G$\cdot$cm$^3\cdot$g$^{-1}$, Model 2: $\mu_{\rm m}=3.88\times
10^{-4}$ G$\cdot$cm$^3\cdot$g$^{-1}$, Model 3: $\mu_{\rm
v}=2.57\times 10^{11}$ G, and Model 4: $\mu_{\rm v}=4.21\times
10^{11}$ G.
According to Eq.(\ref{ppdot}) and the mass-radius relations of
strange stars [numerically calculated, with the state equation
Eq.(\ref{eos}) and the TOV equation], choosing the model
parameters above, we can find values of $P{\dot P}$ for certain
masses, which are represented in Fig.2.
It is found that the $P{\dot P}$ value is limited by either the
maximum mass (models 1 and 3) or by mass-radius relations (models
2 and 4). The maximum $P{\dot P}$ value could be a few $10^{-15}$
s.

In model 1 with $\mu_{\rm m}=5\times 10^{-4}$
G$\cdot$cm$^3\cdot$g$^{-1}$, a pulsar with certain mass
($2M_\odot$, $1.5M_\odot$, and $1M_\odot$, respectively) evolves
along constant $P{\dot P}$ (solid lines in Fig.3).
\begin{figure}
\centerline{\psfig{file=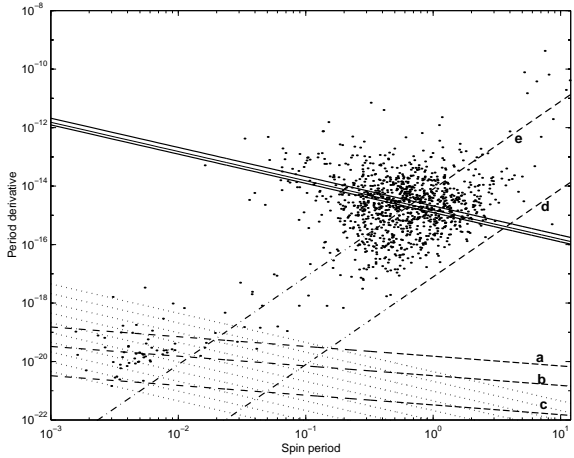,width=3in}}
\caption{The $P{\dot P}$ diagram of pulsars. Three solid lines
from top are for constant $P{\dot P}$ values of pulsars with
masses $2M_\odot$, $1.5M_\odot$, and $1M_\odot$, respectively, in
model 1 ($\mu_{\rm m}=5\times 10^{-4}$
G$\cdot$cm$^3\cdot$g$^{-1}$). Dotted lines are also for constant
$P{\dot P}$, but lower $\mu_{\rm m}$ ($10^{-6}$
G$\cdot$cm$^3\cdot$g$^{-1}$). These ten dotted lines from top are
for pulsars with mass $M_\odot$, $10^{-1}M_\odot$,
$10^{-2}M_\odot$, $10^{-3}M_\odot$, $10^{-4}M_\odot$,
$10^{-5}M_\odot$, $10^{-6}M_\odot$, $10^{-7}M_\odot$,
$10^{-8}M_\odot$, and $10^{-9}M_\odot$, respectively.
The lines labelled {\bf a}, {\bf b}, {\bf c}, {\bf d}, and {\bf e}
are of constant potential drops $\phi$. See the text for the
parameters of these 5 lines.
The pulsar data are downloaded from
http://www.atnf.csiro.au/research/pulsar/psrcat.%
\label{fig2}}
\end{figure}
These three lines pass through almost the middle region of normal
pulsars. It is thus suggestive that the distribution of scattered
points of normal pulsars could be the result of the variation of
$\mu_{\rm m}$, rather than that of pulsar mass. Actually, if the
mass is fixed to be $1.5M_\odot$, most of the normal pulsars are
between $\mu_{\rm m}=5\times 10^{-3}$ and $5\times 10^{-5}$
G$\cdot$cm$^3\cdot$g$^{-1}$.
However, in model 1 with lower $\mu_{\rm m}$ for millisecond
pulsars (e.g., $10^{-6}$ G$\cdot$cm$^3\cdot$g$^{-1}$), the line
with mass $\sim 1M_\odot$ of constant $P{\dot P}$ does {\em not}
pass through the center of millisecond pulsar points. It is
unlikely that the variance of $\mu_{\rm m}$ is responsible for
this unless the $\mu_{\rm m}$ values of those millisecond pulsars
listed in Table 1 are not representative. Therefore, we propose
alternatively that some of the millisecond pulsars could be
low-mass bare strange stars, with mass $\ll M_\odot$.
Similar conclusions can be obtained in the other models (i.e.,
models 2, 3, and 4).

For low-mass bare strange stars, the value $P{\dot P}$ can be
obtained analytically.
The mass ($\leq \sim M_\odot$) of bare strange stars can be well
approximated by \cite{afo86}, because of Eq.(\ref{eos}),
\begin{equation}
M={4\over 3}\pi R^3(4\bar{B}),%
\label{mr}
\end{equation}
where $\bar{B}=(60\sim 110)$ MeV/fm$^3$, i.e., $(1.07\sim
1.96)\times 10^{14}$ g/cm$^3$.
Combing Eq.(\ref{ppdot}) and Eq.(\ref{mr}), one comes to
\begin{equation}
P{\dot P}={320\pi^3\mu_{\rm m}^2\over 9c^3}{\bar B}R,%
\label{ppdotm}
\end{equation}
for models 1 and 2, and
\begin{equation}
P{\dot P}={20\pi^3\mu_{\rm v}^2\over 9c^3}{R\over {\bar B}},%
\label{ppdotv}
\end{equation}
for models 3 and 4 in the low-mass limit.

{\em Lines of constant potential drops in the $P-{\dot P}$
diagram.} The potential drop in the open-field-line region is
essential for pulsar magnetospheric activity.
We adopt only Model 1 in the following indications. From
Eq.(\ref{B}), (\ref{mu}), and (\ref{mr}), one has
\begin{equation}
B={32\pi\over 3}{\bar B}\mu_{\rm m}.%
\label{B-lowmass}
\end{equation}
This shows that the polar fields of homogenously magnetized quark
stars, with certain $\mu_{\rm m}$, of different low masses are
approximately the same. The potential drop between the center and
the edge of a polar cap is
\cite{rs75}%
\begin{equation}
\phi={2\pi^2\over c^2}R^3BP^{-2}.%
\label{phi}
\end{equation}
In case of approximately constant $\mu_{\rm m}$, Eq.(\ref{phi})
can be conveniently expressed as, from Eq.(\ref{B-lowmass}),
\begin{equation}
\phi = {64\pi^3\over 3 c^2}{\bar B}\mu_{\rm m}R^3P^{-2}.%
\label{phi-mu}
\end{equation}
From Eq.(\ref{ppdotm}) and (\ref{phi-mu}), one has
\begin{equation}
P{\dot P}^3={2.026\times 10^6\over c^7}{\bar B}^2\mu_{\rm m}^5\phi.%
\label{phi-mu-const}
\end{equation}
However, if the variance of pulsar masses (or radii) is smaller
than that of $\mu_{\rm m}$, it is better to express potential drop
as, from Eq.(\ref{b-lowmass}) and Eq.(\ref{phi}),
\begin{equation}
\phi = 2\pi\sqrt{{16\pi\over 5c}{\bar B}R^5{\dot P}P^{-3}}, %
\label{phi-mr}
\end{equation}
where Eq.(\ref{mr}) is included.
The lines of constant $\phi$ are drawn in Fig.3, based on both
Eq.(\ref{phi-mu-const}) (dashed lines labelled {\bf a}, {\bf b},
and {\bf c}, with a slop of $-1/3$) and Eq.(\ref{phi-mr})
(dash-dotted lines labelled {\bf d} and {\bf e}, with a slop of
$3$). The parameters for these lines are as following. {\bf a}:
$\phi=10^{12}$ V, $\mu_{\rm m}=10^{-6}$
G$\cdot$cm$^3\cdot$g$^{-1}$; {\bf b}: $\phi=10^{10}$ V, $\mu_{\rm
m}=10^{-6}$ G$\cdot$cm$^3\cdot$g$^{-1}$; {\bf c}: $\phi=10^{12}$
V, $\mu_{\rm m}=10^{-7}$ G$\cdot$cm$^3\cdot$g$^{-1}$; {\bf d}:
$\phi=10^{12}$ V, $R=10$ km; and {\bf e}: $\phi=10^{11}$ V, $R=1$
km.

Pair production mechanisms are essential for pulsar radio
emission. A pulsar is called to be ``death'' if the pair
production condition can not be satisfied. A general review of
understanding radio pulsar death lines can be found in
\cite{zhang03}.
Although a real deathline depends upon the dynamics of detail pair
and photon production, the deathline can also be conventionally
taken as a line of constant potential drop $\phi$.
It is found in Fig.3 that the slop of constant $\phi$ is $-1/3$
(or $3$) if the scattering distribution of pulsar points in
$P-{\dot P}$ diagram is due to different masses (or polar field
$B$) but with constant $\mu_{\rm m}$ (or mass or radius).
The deathline slop may be expected to be between $-1/3$ and $3$ if
the distributions of mass and polar field are combined.

\subsection{Accretion-dominated spindown}

The physical process of accretion onto rotating pulsar-like stars
with strong magnetic fields is very complex but is essential to
know the astrophysical appearance of the stars (e.g., the
variation of X-ray flux, the evolutional tracks, etc.), which is
still not be understood well enough \cite{lip92}. Nevertheless, it
is possible and useful to describe the accretion
semi-quantitatively.

For an accretion scenario in which the effect of kinematic energy
of accreted matter at infinite distance is negligible (such as the
case of supernova fall-back accretion), three typical radii are
involved. The radius of light cylinder of a spinning star with
period $P$ is
\begin{equation}
r_{\rm l}={cP\over 2\pi}=4.8\times 10^9P~{\rm cm}. %
\label{rl}
\end{equation}
If all of the accretion material is beyond the cylinder, the star
and the accretion matter could evolve independently.
The magnetospheric radius, defined by equating the kinematic
energy density of free-fall particles with the magnetic one
$B^2/(8\pi)$, is
\begin{equation}
r_{\rm m}=({B^2R^6\over {\dot M}\sqrt{2GM}})^{2/7}, %
\label{rm}
\end{equation}
where $\dot M$ is the accretion rate. In the low-mass limit of
bare strange stars, considering the mass-radius relation of
Eq.(\ref{mr}), Eq.(\ref{rm}) becomes
\begin{equation}
\begin{array}{lll}
r_{\rm m}&=&({3\over 32\pi G})^{1/7} {\bar B}^{-1/7}
B^{4/7}R^{9/7}{\dot M}^{-2/7}\\
 & = &0.064{\bar B}_{60}^{-1/7}B^{4/7}R^{9/7}{\dot M}^{-2/7},
\end{array}
\label{rm-lowmass}
\end{equation}
where the bag constant ${\bar B}={\bar B}_{60}\times 60$
MeV/fm$^3$. If a star is homogenously magnetized per unit mass
(i.e., in Models 1 and 2), according to Eq.(\ref{B-lowmass}), one
has from Eq.(\ref{rm-lowmass})
\begin{equation}
\begin{array}{lll}
r_{\rm m}&=&({32^3\pi^3\over 27G})^{1/7} {\bar B}^{3/7}\mu_{\rm
m}^{4/7}R^{9/7}{\dot M}^{-2/7}\\
 & = &4.9\times 10^7{\bar B}_{60}^{3/7}\mu_{\rm
m}^{4/7}R^{9/7}{\dot M}^{-2/7}.
\end{array}
\label{rm-model-1-2}
\end{equation}
Due to the strong magnetic fields around a spinning star, matter
is forced to co-rotate, and both gravitational and centrifugal
forces work. At the so-called corotating radius $r_{\rm c}$, these
two forces are balanced,
\begin{equation}
r_{\rm c}=({GM\over 4\pi^2})^{1/3}P^{2/3}=1.2\times 10^{-3}M^{1/3}P^{2/3}. %
\label{rc}
\end{equation}
In the low-mass limit, one has from Eq.(\ref{mr}) and
Eq.(\ref{rc})
\begin{equation}
\begin{array}{lll}
r_{\rm c}&=&({4G\over 3\pi})^{1/3} {\bar B}^{1/3}
RP^{2/3}\\
 & = &145{\bar B}_{60}^{1/3}RP^{2/3}.
\end{array}
\label{rc-lowmass}
\end{equation}

In another case, in which the kinematic energy at infinity is {\em
not} zero (i.e., ISM or stellar wind accretion), besides those
three radii, an additional one is the accretion radius $r_{\rm
a}$, at which the total energy (kinematic and gravitational ones)
is zero,
\begin{equation}
r_{\rm a}=2GMV_\infty^{-2}, %
\label{ra}
\end{equation}
where $V_\infty$ is the relative velocity of the star to the
surrounding media. The motion of matter only at a radius $<r_{\rm
a}$ could be affected by gravity, and the mass capture rate is
then
\begin{equation}
{\dot M}_{\rm c}=\pi r_{\rm a}^2 \rho V_\infty =4\pi G^2M^2\rho V_\infty^{-3}, %
\label{mdot}
\end{equation}
where $\rho$ is the density of diffusion material.

Due to the centrifugal inhibition, since the radius of matter
nearest to the star could be $r_{\rm m}$, massive accretion onto
stellar surface is impossible when $r_{\rm m}>r_{\rm c}$. This is
the so-called {\em supersonic} propeller spindown phase. A star
spins down to the equilibrium period $P_{\rm eq}$, defined by
$r_{\rm m}=r_{\rm c}$. In the low-mass limit, one has, from
Eq.(\ref{rm-lowmass}) and Eq(\ref{rc-lowmass}),
\begin{equation}
P_{\rm eq}=0.72 G^{-5/7}{\bar B}^{-5/7}B^{6/7}R^{3/7}{\dot M}^{-3/7}, %
\label{Peq}
\end{equation}
or assuming a homogenous magnetic momentum per unit mass, from
Eq.(\ref{rm-model-1-2}) and Eq(\ref{rc-lowmass}),
\begin{equation}
P_{\rm eq}=15 G^{-5/7}{\bar B}^{1/7}\mu_{\rm m}^{6/7}R^{3/7}{\dot M}^{-3/7}. %
\label{Peq-mum}
\end{equation}
However, accretion with rate $\dot M$ onto the stellar surface is
not possible, although the centrifugal barrier is not effective
when $P>P_{\rm eq}$, until the star spins down to a so-called
break period \cite{dfp79,dp81,ik03},
\begin{equation}
\begin{array}{lll}
P_{\rm br}&=&60 \mu_{30}^{16/21}{\dot M}_{15}^{-5/7}M_1^{-4/21}~{\rm s}\\
 & = & 36 {\bar B}_{60}^{-4/21}B_{12}^{16/21}{\dot M}_{15}^{-5/7}R_6^{12/7}~{\rm
 s}\\
 & = & 0.49 {\bar B}_{60}^{12/21}\mu_{m-6}^{16/21}{\dot M}_{15}^{-5/7}R_6^{12/7}~{\rm
 s},
\end{array}
\label{Pbr-mum}
\end{equation}
in the low-mass limit, where Eq.(\ref{B}), Eq.(\ref{mr}), and
Eq.(\ref{B-lowmass}) have been included, and the convention
$Q=10^nQ_n$ has been adopted.
Pulsars with periods between $P_{\rm eq}$ and $P_{\rm br}$ do
still spin down. This phase is called as {\em subsonic} propeller.
Only a negligible amount of accretion matter can penetrate into
the magnetosphere (onto the stellar surface) during both the
supersonic and subsonic propeller phases \cite{ik03}, and the
expected accretion luminosity is thus very low.

How to determine quantitatively the spindown rate when a pulsar is
in those two propeller phases? No certain answer known hitherto
for the propeller torques, even not to be certain about the
accretion configuration (disk or sphere). Nonetheless, if the
spinup effect of matter accreted onto stellar surface is
neglected, the spindown rate can be estimated according to the
conservation laws of angular-momentum and/or rotational-energy
\cite{dfp79}.
The escape velocity at radius $r$ is $\sqrt{2GM/r}$. Approximating
the stellar angular-momentum loss rate $-2\pi I{\dot P}/P^2$ to
that of accretion material near $r_{\rm m}$ (based on the
angular-momentum conservation), we have
\begin{equation}
\begin{array}{lll}
{\dot P}_{\rm J}&=&{\sqrt{G}\over \sqrt{2}\pi} M^{1/2}I^{-1}{\dot
M}r_{\rm
m}^{1/2}P^2\\
 & = & {5\sqrt{6G}\over 8\pi^{3/2}} {\bar B}^{-1/2}R^{-7/2}{\dot M}r_{\rm
m}^{1/2}P^2,%
\end{array}
\label{Pdot-angular}
\end{equation}
where Eq.(\ref{mr}) is introduced.
However, in case of energy conservation, $I\Omega{\dot
\Omega}=-GM{\dot M}/r_{\rm m}$, one comes to
\begin{equation}
\begin{array}{lll}
{\dot P}_{\rm E}&=&{G\over 4\pi^2} MI^{-1}{\dot M}r_{\rm
m}^{-1}P^3\\
 & = & {5G\over 8\pi^2} R^{-2}{\dot M}r_{\rm m}^{-1}P^3.%
\end{array}
\label{Pdot-energy}
\end{equation}

\section{The case of 1E 1207.4-5209}

The radio-quiet central compact object in the supernova remnant
PKS 1209-51/52, 1E 1207.4-5209, is a unique pulsar-like star which
is worth noting since we known much information about it: the
rotating period $P=0.424$ s, the cyclotron energy $E_{\rm
cyc}=0.7$ keV \cite{big03}, the age $T\sim 7$ kys estimated from
the remnant, with an uncertainty of a factor of 3 \cite{rog88},
the timing properties \cite{zps04}, and the thermal X-ray spectrum
of long-time observations \cite{luca04}.
The distance to the remnant is $d=1.3\sim 3.9$ kpc, the X-ray flux
in a range of 0.4-8 keV is $2.3\times 10^{-12}$ erg cm$^{-2}$
s$^{-1}$, and the corresponding X-ray luminosity is then
$L=(0.47\sim 4.2)\times 10^{33}$ erg/s \cite{pavlov04}.
However, the more we observe, the knottier astrophysicists can
have a model to understand its nature.

Two issues are addressed at first. One is about its absorption
lines.
The lines at 0.7, 1.4, and 2.1 keV (and possibly 2.8 keV) are
identified, which are phase-dependent \cite{mer02}. These imply a
cyclotron-origin of the features \cite{big03,xwq03}, although this
possibility was considered to be unlikely when discovered by {\em
Chandra} \cite{san02}.
However, there are still two questions relevant to this issue.

(1) Where does the absorption form (near the stellar surface or in
the magnetosphere)?
An e$^\pm$ plasma surrounding a magnetized neutron star,
maintained by the cyclotron-resonance process, was suggested to
prevent a direct detection of the stellar surface in X-ray band,
the existence of which seems to explain the age dependance of the
effective radiating area \cite{rudm03}. The cyclotron lines would
thus form at a height where resonant scattering occurs. In the
regime, all neutron stars with high B-fields should present
cyclotron absorption in their thermal X-ray spectra; but this
conflicts with the observations. In addition, the physics of the
plasma is still not well studied, and its density and stability
are not sure.
An alternative and intuitive suggestion is that the line forming
region is near the stellar surface. In this case, we may need bare
quark surface, with an electron layer of density $\sim 10^{32}$
cm$^{-3}$ and thickness of a few thousands of fermis, in order to
explain those absorption dips. Due to the degeneracy of electrons,
only electrons near the surface of fermi-sea can be exited to
higher levels, the the number of which is energy-dependent. For
instance, the number of electrons which resonantly scatter photons
with energy $\sim 1.4$ keV could be about the double of that with
$\sim 0.7$ keV. The number of electrons, which are responsible to
cyclotron-resonant of photons with higher energy, is therefore
larger although absorption cross-section is smaller. Another
factor, which may also favor more electrons to absorb higher
energy photons, could be of radiative transfer process (e.g., a
certain layer might be optically thick at $\sim 0.7$ keV, but
optically thin at $\sim 2.1$ ekV), but a detail consideration on
this is necessary in the future. We conclude then that it is
reliable to assume a surface origin for the cyclotron-resonant
lines.

(2) Is the cyclotron resonant in terms of electrons or protons?
The fundamental electron cyclotron resonant lies at $\Delta E_{\rm
e}=11.6B_{12}\sqrt{1-r_{\rm s}/R}$ keV, while the proton one at
$\Delta E_{\rm p}=6.3B_{12}\sqrt{1-r_{\rm s}/R}$ eV, where $r_{\rm
s}\equiv 2GM/c^2$ is the Schwartzchild radius.
For a star with $10^6$ cm and $1M_\odot$, the factor
$\sqrt{1-r_{\rm s}/R}=0.84$. In case of lower mass strange stars,
$r_{\rm s}/R\sim R^2$ is smaller, and the factor is closer to 1.
We thus just approximate the factor to be 1, the field is then
$B=6\times 10^{10}$ G in terms of electrons, $B=10^{14}$ G of
protons.
If the lines are proton-originated, there is still two scenarios.
One is that the multipole fields have strength $B_{\rm m}\sim
10^{14}$ G, but the global dipole field $B_{\rm p}\sim 3\times
10^{12}$ G is much smaller in order to reconcile the spindown rate
expected from Eq.(\ref{dotO}). This means that the stellar surface
is full of flux loops with typical length $l_{\rm loop}$, which
can be estimated to be \cite{TD93},
\begin{equation}
l_{\rm loop}\sim \sqrt{B_{\rm p}\over B_{\rm m}}R
\sim 10^5R_6~{\rm cm}. %
\label{loop}
\end{equation}
The maximum release energy due to magnetic reconnection could be
$\sim (B_{\rm m}^2/8\pi)l_{\rm loop}^3\sim 10^{42}$ erg. If the
dynamical instability takes place at a short timescale of $\sim 1$
s, as observed in soft $\gamma$-ray repeaters, bursts with $\sim
10^{42}$ erg/s might have been detected in 1E 1207.4-5209; but we
do not.
Another scenario is that the dipole field of 1E 1207.4-5209 is
$\sim 10^{14}$ G, but the accreted material onto the stellar
surface contributes a positive angular momentum. The magnetodipole
radiation should spin down the object at a rate ${\dot P}=2\times
10^{-10}$ s/s, based on Eq.(\ref{b}), which is much larger than
observed ($\sim 10^{-14}$ s/s). This discrepancy might be
circumvented if accreted matter contributes a positive momentum.
Yet, this can only be possibly when $P>P_{\rm br}$, which results
in an accretion rate, according to Eq.(\ref{Pbr-mum}), ${\dot
M}>0.9 \times 10^{15}$ g/s and an X-ray luminosity $L_{\rm
x}>10^{35}$ erg/s, to be much larger than observed $L\sim 10^{33}$
erg/s, for typical neutron star parameters.
In both pictures, however, a strange thing is: why does not this
star with magnetar-field show magnetar-activity (e.g., the much
higher luminosity $\sim 10^{34-35}$ erg/s of persistent X-ray
emission, observed in anomalous X-ray pulsars and soft
$\gamma$-ray repeaters)?
In addition, there is still no idea to answer why the feature
strength is similar at $\sim 0.7$ and $\sim 1.4$ keV (and even the
appearance of a line at $\sim 2.1$ keV), due to the high
mass-energy of protons ($\sim 10^3$ times that of electrons), as
discussed in the case of SGR 1806-20 \cite{xwq03}.
We therefore tend to suggest an electron-cyclotron-origin of the
absorption features.

Another issue is about its radius detected.
In principle, one can obtain a radius of a distant object by
detecting its spectrum (fitting the spectrum gives out a
temperature $T$ {\em if} a Plankian spectrum approximation is good
enough) and flux $F$, through $F=\sigma T^4\cdot R^2/d^2$ ($R$ is
radiation radius, not the coordinate radius $R_{\rm coord}$ in the
Schwartzschild metric. $R=R_{\rm coord}/\sqrt{1-r_{\rm s}/R_{\rm
coord}}$ if the spectrum is Plankian), if the distance $d$ is
measured by other methods (e.g., parallax). However, we are not
sure if the thermal spectrum of 1E 1207.4-5209 is really Plankian,
and we do not known the distance neither.
Nonetheless, since the spectrum depends also on the ISM absorption
(the neutral hydrogen density is supposed to be known), one may
fit the spectrum by free parameters $T$, $d$, and $R$. Note that
radii determined in this way are highly uncertain.
A radius of $\sim 1$ km was obtained in single-blackbody models by
{\em ROSAT} \cite{mer96} and {\em ASCA} \cite{vas97} observations,
whereas a 10 km-radius was suggested in an atmosphere model of
light-elements by \cite{zpt98}.  An {\em XMM-Newton} observation
yields recently two-blackbody radii fitted: $R=0.8$ km and $4.6$
km for hotter and cooler components, respectively \cite{luca04}.
The possibility, that 1E 1207.4-5209 may have a radius to be much
smaller than 10 km of conventional neutron stars, is thus not
ruled out.

Now we turn to a low-mass bare strange star model for 1E
1207.4-5209.
Besides the absorption features, the most outstanding nature, that
makes the pulsar be puzzling enough, should be its timing
behavior: it does not spin down stably but seems to {\em spin up}
occasionally. Furthermore, two or more probability
peak-frequencies are identified during each of the five
observations \cite{zps04}. Pulsar glitches and Doppler shift in a
binary system were proposed for the ``spinup''
\cite{zps04,luca04}, but the multi-frequency distributions are
still not well understood.
An alternative model proposed here is that the pulsar is a
low-mass bare strange star which is at a critical point of its
subsonic propeller phase, $P\sim P_{\rm br}$.

Steady accretion onto a magnetized and spinning star is possible
only if $P>P_{\rm br}$, when magnetohydrodynamic (e.g.,
Rayleigh-Taylor and Kelvin-Helmholtz) instabilities occur at the
magnetospheric boundary \cite{al76,el84}. Before the onset of the
instabilities, accretion plasma can only penetrate into the
magnetosphere by diffusion, with a rate much smaller than $\dot M$
\cite{ik03}. Propeller torque may spin down a star to a period of
$P\ga P_{\rm br}$. At this period, steady accretion does not occur
until the density just outside the boundary increases to a
critical density of $\rho_{\rm crit}\sim 7\rho_{\rm m}$ when the
Rayleigh-Taylor instability happens \cite{el84}, where $\rho_{\rm
m}\sim {\dot M}/(4\pi r_{\rm m}^2\sqrt{2GM/r_{\rm m}})$. The star
should then be spun up by steady accretion torque\footnote{
In case of wind--fed accretion in binary systems, the star will
spin up ({\em or} down) if the accreted material has a positive
(negative) angular momentum. Whereas, in case of fallback disk
accretion or ISM-fed accretion, the momentum of accreted matter
should be positive, and the steady-accretion-induced torque leads
the star to spin up. We are not considering accretion in a binary
system here.}.
However, increasing the spin-frequency may dissatisfy the
necessary condition for steady accretion, $P>P_{\rm br}$. The star
will then return back to a subsonic propeller phase when the
decreasing period is low enough, and it spins down again.
We could therefore expect an erratic timing behavior when a pulsar
is at this critical phase, $P\sim P_{\rm br}$. The object 1E
1207.4-5209 could be an ideal laboratory for us to study the
detail physics in such an accretion stage.

For 1E 1207.4-5209, setting $P_{\rm br}=0.424$ s, $B_{12}=0.06$,
one has $\mu_{\rm m}=9.1\times 10^{-6}{\bar B}_{110}^{-1}$
G$\cdot$cm$^3\cdot$g$^{-1}$ from Eq.(\ref{B-lowmass}), which is
close to that values of millisecond pulsars (Table 1). The
accretion rate reads, from Eq.(\ref{Pbr-mum}),
\begin{equation}
{\dot M}=10^{14} {\bar B}_{60}^{-4/15}R_5^{12/5}~~{\rm g/s},
\label{mdot1207}
\end{equation}
and the magnetospheric radius is then, from Eq.(\ref{rm-lowmass}),
\begin{equation}
r_{\rm m}=2.5\times 10^7{\bar B}_{60}^{-1/15}R_5^{3/5}~~{\rm cm}.
\label{rm1207}
\end{equation}
As for the instantaneous spindown rate in the subsonic propeller
phase, we apply Eq.(\ref{Pdot-angular}) and Eq.(\ref{Pdot-energy})
for estimation, according to the momentum and energy
conservations, respectively.
One has then,
\begin{equation}
\left\{
\begin{array}{lll}
{\dot P}_{\rm J}&=&1.9\times 10^{-12}{\bar B}_{60}^{-4/5}R_5^{-4/5},\\
{\dot P}_{\rm E}& = & 1.3\times 10^{-13}{\bar B}_{60}^{-1/5}R_5^{-1/5}.%
\end{array}
\right. %
\label{Pdot1207}
\end{equation}
These results imply that the instantaneous period increase could
be one or two orders larger than the averaged one ($\sim 10^{-14}$
s/s) if $R_5\sim 1$. A precision measurement of instantaneous
${\dot P}_{\rm inst}$ may tell us the actual radius.
In fact, possible {\em spinup} has been noted in the {\em
XMM-Newton} observations of August 2002, with an instantaneous
period decrease rate ${\dot P}_{\rm inst}=-(3\sim 6)\times
10^{-14}$, whereas {\em spindown} in {\em Chandra} observations of
June 2003, with ${\dot P}_{\rm inst}\sim 2\times 10^{-13}$ to be
much larger than the averaged one \cite{zps04}. The conjectured
radius is then likely to be $\sim 1$ km, based on this
instantaneous rate and the spindown rule of Eq.(\ref{Pdot-energy}
), and 1E 1207.4-5209 could be a low-mass bare strange star
($M\sim 10^{-3}M_\odot$).
Such a high accretion rate ($\sim 10^{14}$ g/s) can certainly not
of the capture by the star from interstellar medium matter, but
could be due to a fallback flow, sine the age is relatively young
($\sim 7$ kys).
The age discrepancy between the SNR age and that\footnote{%
See Eq.(\ref{T}), for the case of $P_0\ll 0.424$ s.
} %
of $P/(2{\dot P})$ is not surprising since the pulsar is not
dominantly rotation-powered. It depends on the detail physics of
magnetospheric boundary interactions to solve the age problem for
this pulsar.

The accretion X-ray luminosity, for steady accretion with rate of
$\dot M$, could be $\sim GM{\dot M}/R\sim 10^{32}$ erg/s, which is
comparable with the observed $L\sim 10^{33}$ erg/s. This hints
that the X-ray radiation of 1E 1207.4-5209 could be powered by
both of cooling and accretion. Note that, from Eq.(\ref{dotO}),
the energy-loss rate due to magnetodipole radiation from such a
low-mass star is only $\sim 10^{24}$ erg/s, to be much smaller
than the X-ray luminosity $L$. Therefore, such pulsar-like stars
have negligible magnetospheric activities, and can be observed
only if they are near and young enough.

The models proposed, including this strange-star model and that of
\cite{zps04}, can be tested by future timing observations. The
Doppler shift model should be ruled out if future pulses are not
arrive at the time expected in the model. If random spin nature
can be confirmed as a general nature in more precise observations,
we may tend to suggest $P\sim P_{\rm br}$ for 1E 1207.4-5209,
since glitching pulsars usually spin stably except during
glitches.

\section{Other candidates of low-mass strange stars}

A first candidate pulsar of low-mass strange stars was suggested
in 2001 \cite{xxw01}: the fastest rotating millisecond pulsar PSR
1937+21 ($P=1.558\times 10^{-3}$ s, ${\dot P}=1.051\times
10^{-19}$ s/s).
In order to explain its polarization behavior of radio pulses and
the integrated profile (pulse widths of main-pulse and
inter-pulse, and the separation between them), this pulsar is
supposed to have mass $<0.2M_\odot$ and radius $<1$ km.
The polar magnetic field is $8.2\times 10^8$ G based on
Eq.(\ref{b}) for conventional neutron stars, but could be
$2.2\times 10^9{\bar B}_{60}^{1/2}$ G if the radius $R=1$ km and
Eq.(\ref{b-lowmass}) and Eq.(\ref{mr}) are applied.

The low mass may actually favor a high spin frequency $\Omega$
during the birth of a bare strange star. The defined Kepler
frequency of such stars could be approximately a constant,
\begin{equation}
\Omega_{\rm k}=\sqrt{GM\over R^3}=1.1\times 10^4{\bar
B}_{60}^{1/2}~~{\rm s^{-1}},
\label{kepler}
\end{equation}
with a prefactor of $\sim 0.65$ at most for $M\sim M_\odot$ and
$R\sim 10^6$ cm \cite{glen00}.
The initial rotation periods of strange stars are limited by the
gravitational radiation due to $r$-mode instability \cite{mad98}.
At this early stage, the stars are very hot, with temperatures of
a few $10$ MeV, and we may expect a fluid state of quark matter,
without color superconductivity.
The critical $\Omega$ satisfies the equation
\begin{equation}
{1\over \tau_{\rm gw}}+{1\over \tau_{\rm sv}}+{1\over \tau_{\rm
bv}}=0,
\label{r-mode}
\end{equation}
where the growth timescale for the instability (the negative sigh
indicates that the model is unstable) is estimated to be,
\begin{equation}
\tau_{\rm gw}=-3.85\times 10^{81}\Omega^{-6}M^{-1}R^{-4},
\label{taugw}
\end{equation}
and $\tau_{\rm sv}$ and $\tau_{\rm bv}$ are the dissipation
timescales due to shear and bulk viscosities, respectively,
\begin{equation}
\left\{
\begin{array}{lll}
\tau_{\rm sv}& = & 1.85\times 10^{-9}\alpha_{\rm s}^{5/3}M^{-5/9}R^{11/3}T^{5/3},\\
\tau_{\rm bv}& = & 5.75\times 10^{-2}m_{100}^{-4}\Omega^2R^2T^{-2},%
\end{array}
\right. %
\label{taudiss}
\end{equation}
and $\alpha_{\rm s}$ the coupling constant of strong interaction,
$T$ the temperature, $m_{100}$ the strange quark mass in 100 MeV.

\begin{figure}
\centerline{\psfig{file=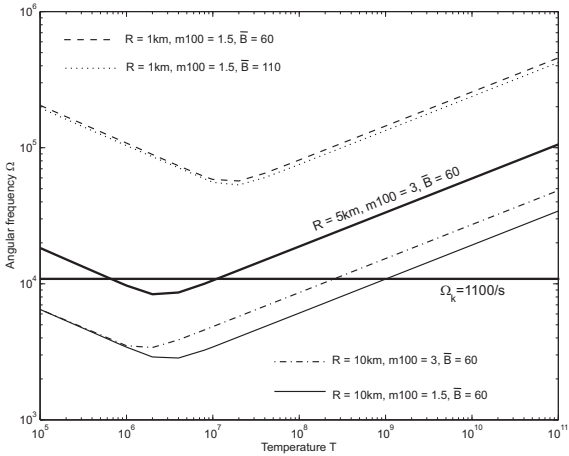,width=3in}}
\caption{Temperature dependence of the maximum angular frequency
in rotating bare strange stars, due to the gravitational
instability in the $r$-modes. The coupling constant $\alpha_{\rm
s}=0.1$. Other parameters are shown for each types of lines. The
bag constant $\bar B$ is in unit of MeV/fm$^3$.
\label{rmode}}
\end{figure}
The calculated results, based on Eq.(\ref{r-mode}), are shown in
Fig.4. It is found that low-mass bare strange stars can rotate
very fast, even faster than the Kepler frequency (Note: the
surface matter is not broken at super-Kepler frequency, due to the
self-bounding of quark matter), and one would then not be
surprising that the fastest rotating pulsar could be a low-mass
bare strange star.
However, though it needs advanced technique of data collection and
analysis to detect sub-millisecond radio pulsar, we have not find
one yet \cite{esb01,han04}. This negative result could be due to:
(1) The dynamical process does not result in a sub-millisecond
rotator; (2) No magnetospheric activity exists for very low-mass
strange stars whose $\dot P$ is very small, since the potential
drop is not high enough to trigger pair production (see \S3.1). In
the later case, a nearby sub-millisecond radio pulsar could be
found by X-ray observations since a hotspot, powered by rotation
and/or accretion, may form on the stellar surface.

RX J1856.5-3754 is another candidate. RX J1856.5-3754 could be
low-massive, based on the X-ray spectrum, but the main puzzle is
the origin of its optical radiation, the intensity of which is
about 7 times that extrapolated from the Rayleigh-Jeans law of
X-ray spectrum \cite{bur03}.
If RX J1856.5-3754 is a spinning magnetized star, its
magnetosphere could be surrounded by a spherically quasi-static
atmosphere, in which the plasma temperature is of the order of the
free-fall temperature \cite{lip92,ik03},
\begin{equation}
T_{\rm ff} = {GMm_{\rm p}\over kr_{\rm m}},
\label{Tff}
\end{equation}
where $m_{\rm p}$ the proton mass and $k$ the Boltzmann constant.
The dissipation of stellar rotation energy, as well as the
gravitational energy of accreted matter, may heat the envelop,
which could be responsible for the UV-optical emission. The
soft-component-fitted parameters could thus be of this quasistatic
envelop, with temperature $<33$ eV and radius $>17$ km. Assuming
$T_{\rm ff}<33$ eV and $r_{\rm m}>17$ km, one has low
limit\footnote{
For blackbody radiation, $r_{\rm m}T^2$ is a constant. Accordint
to Eq.(\ref{Tff}), one come to $M\sim r_{\rm m}T\sim (r_{\rm
m}T^2)T^{-1}$.
} %
of the stellar mass $M>4\times 10^{-7}M_\odot$, or radius $R>0.1$
km. While the hard-component-fitted stellar radius is $R=4.4$ km
\cite{bur03}. One can also infer an accretion rate ${\dot M}\sim
4\times 10^{10}$ g/s, from Eq.(\ref{rm-model-1-2}) for $\mu_{\rm
m}=10^{-6}$ G$\cdot {\rm cm}^3\cdot$g$^{-1}$. The X-ray luminosity
due to accretion is then $> 3\times 10^{26}$ erg/s. This model for
soft component could be tested by more observations in sub-mm
bands, besides in optical and UV bands, since the quasi-static
atmosphere could also be effective in radiating infrared photons
if it is dusty.

Strange quark matter with mass $\ll M_\odot$ could be ejected by a
massive strange star ($\sim M_\odot$) during its birth or by
collision of two strange stars, and such low-mass matter may
explain a few astrophysical phenomena \cite{xw03}: the planets
around pulsars could be quark matter with mass $\sim 10^{23-28}$
g, while very low mass strange quark matter (called as
strangelets) with baryon numbers of $\sim 10^9$ may be the nature
of ultra-high energy cosmic rays beyond the GZK cutoff.
The bursts of soft $\gamma$-ray repeaters could be due to either
the starquake-induced magnetic reconnection or the collision
between a strange planet and solar-mass bare strange star. The
collision chance would be not low if both objects form during a
same supernova, or in a same binary system.
Some of the transient unidentified EGRET sources \cite{wal00} may
represent such collision events, the gravitational energy release
of which is
\begin{equation}
E_{\rm g} \sim {GM_{\rm a}M_{\rm b}\over (R_{\rm a}^3+R_{\rm
b}^3)^{1/3}} = 2\times 10^{23}{\bar B}_{60}^2{R_{\rm a}^3R_{\rm
b}^3\over (R_{\rm a}^3+R_{\rm b}^3)^{1/3}}~{\rm ergs},
\label{Eg}
\end{equation}
where ``a'' and ``b'' denotes two objects of strange quark matter.
The released energy could be $\sim 10^{45}$ erg if $R_{\rm a}\sim
10^5$ cm and $R_{\rm b}\sim 10^4$ cm. This strong (color)
interaction may result in photon emission by various hadron
process (e.g., hadron annihilation), with energy $\ga 100$ MeV
(the EGRET telescope covers an energy range from 30 MeV to over 20
GeV), and could be another way to produce strangelets.

Merging quark stars, rather than neutron stars \cite{elps89}, may
result in cosmic $\gamma-$ray bursts observed (GRBs), which could
help to eliminate the baryon load problem.
The released energy is $\sim 10^{53}$~ergs during the collision of
two quark stars with $\sim 10^6$ cm. The residual body should be
expected to rapidly rotate, and such a high spin frequency may
result in a beaming pattern of emission.
A fireball with low-baryon contamination in this color interaction
event may favor the emission of photons and neutrinos with high
energy.
Alternatively, rapid rotating quark stars being residual via
hypernovae could also be possible as the central engines. Only
non-baryonic particles (e.g., photons, neutrinos, and $e^\pm$) can
massively radiated from the quark surfaces, since the stellar
temperature is initially very high ($>10$ MeV). A fireball forms
then, and the Usov mechanism \cite{usov98} of pair production in
strong electric field near quark surface may play an important
role at that time. The rotation may cause the very irregular
lightcurves.

\section{The origin of low-mass strange stars}

This is a real problem which is difficult to answer with certainty
now. Several scenarios and arguments relevant are suggested in
this sections, though, in this paper, we focus on proposing
low-mass strange star candidates and trying to attract one's
attention to such stars neglected previously.

1. {\em The origin of millisecond pulsars}.
An open debate on this issue took place in 1996 \cite{Bha96}.
Millisecond pulsars are recycled ones in low-mass X-ray binaries,
the magnetic fields of which decay (by, e.g., enhanced Ohmic
dissipation, diamagnetic screening effect, etc.) during accretion
process, according to the standard model.
After years of searching for coherent millisecond X-ray
pulsations, five accretion-driven millisecond pulsars have been
discovered since 1998 \cite{Wijn04}, which are important to test
the model.
Besides the old problems (e.g., birthrate, millisecond pulsars
with planets, etc.), new puzzling issues are raised in the
standard arguments \cite{rfs04}.

However, the puzzles may disappear if low-mass bare strange stars,
with rapidly spinning, result directly from accretion-induced
collapse (AIC) of white dwarfs, but could be covered by normal
matter if they had high-accretion phases in the later evolutions
(while the core collapses may produce only normal pulsars with
mass $\sim M_\odot$ and radius $\sim 10^6$ cm). Normal neutron
stars created by AIC had been investigated with great efforts
\cite{fryer99}.
A detonation wave, separating nuclear matter and quark matter,
should form inside white dwarfs, and propagate outwards, if quark
stars are born via AIC.
Though the possibility of forming a massive strange star with
$\sim M_\odot$ by AIC might not be ruled out, the reason that AIC
produces low-mass strange star might be simple: {\em The density
and temperature in an accreting white dwarf may not be so high
that the detonation flame reaches near the stellar surface}. A
low-mass quark star could form if the detonation surface is far
below the stellar surface.

The newborn low-mass bare strange stars could rotate very fast,
even to be super-Keplerian (Fig.4), and would spin down if the
accretion rates are not very high.
Unless the mass is smaller than $(0.1\sim 0.3)M_\odot$, the
thickness of crust could be negligible\footnote{
The crust thickness could be much smaller than previous
calculation for static and cold cases due to a high penetration
rate during hot bursts.
} %
\cite{xu03a}, and the maximum X-ray luminosity at the break period
[${\dot M}=2.3\times 10^{16}P_{\rm br-3}^{-7/5}{\bar
B}_{60}^{12/15}\mu_{\rm m-6}^{16/15}R_5^{12/5}$ g/s from
Eq.(\ref{Pbr-mum})] is approximately,
\begin{equation}
L_{\rm x} = {GM{\dot M}\over R} \sim 2.8\times 10^{34}P_{\rm
br-3}^{-7/5}{\bar B}_{60}^{9/5}\mu_{\rm
m-6}^{16/15}R_5^{22/5}~~{\rm erg/s}.
\label{Lx}
\end{equation}
A strange star with low mass may explain the low time-averaged
accretion luminosity in bursting millisecond X-ray pulsars since
$L_{\rm x}\propto R^{4.4}$, if it is assumed that bursting X-ray
pulsars are in a critical stage of $P\sim P_{\rm br}$.
It is worth noting that, in this stage, the real
accretion-luminosity could be much lower than $L_{\rm x}$
presented in Eq.(\ref{Lx}) because only part of the accretion
matter with rate $\dot M$ could bombard directly the stellar
surfaces if $P<P_{\rm br}$.

It is generally suggested that, during an iron-core collapse
supernova, the gravitation-released energy $\mathcal{E}_{\rm
g}\sim 10^{53}$ erg is almost in the form of neutrinos, $\sim
10^{-2}$ of which is transformed into the kinetic energy of the
outgoing shock and $\sim 10^{-4}$ of which contributes to the
photon radiation. However, this idea is not so successful in
modern supernova simulations \cite{janka03,lieb04}, since the
neutrino luminosity could not be large enough for a successful
explosion even in the models with the inclusion of convection
below the neutrinosphere (2D-calculations).
We note that the bare quark surfaces may be {\em essential} for
successful explosions of both types of iron-core collapse and AIC.
The reason is that, because of the color binding, the photon
luminosity of a quark surface is not limited by the Eddington
limit. It is possible that the prompt reverse shock could be
revived by photons, rather than neutrinos.
A hot quark surface, with temperature\footnote{
This is estimated by only gravitational energy release. The
temperature of low-mass bare strange stars could also be
$>10^{11}$ K, since each baryon would release $10 \sim 100$ MeV
during the quark phase transition.
} %
$T>10^{11}$ K, of a newborn strange star\footnote{
The thermal conductivity of quark matter is much larger than that
of normal matter in proto-neutron star crusts. The surface
temperature of proto-neutron stars can not be so high, otherwise a
significant amount of stellar matter should be expelled as wind
(the neutron star mass might then be very small).
} %
will radiate photons at a rate of
\begin{equation}
{\dot E}_{\rm p}>4\pi R^2\sigma T^4\sim 7\times
10^{50}R_5^2T_{11}^4~~{\rm erg/s},
\label{dotEp}
\end{equation}
while the Thomson-scattering-induced Eddington luminosity is only
\begin{equation}
L_{\rm Edd}={64\pi^2cGm_{\rm p}\over 3\sigma_{\rm T}}{\bar
B}R^3\sim 10^{35}{\bar B}_{60}R_5^3~~{\rm erg/s}.
\label{LEdd}
\end{equation}
This means that the photon emissivity may play an important role
in both types of supernova explosions (i.e., for the birth of
solar-mass as well as low-mass bare strange stars).


Strange stars born via this way are certainly bare, since any
normal matter can not survive from the strong photon bursts. A
high fall-back accretion may not be possible due to massive
ejecta, rapid rotation, and strong magnetic fields, and such stars
could keep to be bare as long as the accretion rates are not very
high, since accreted matter with low rates could penetrate the
Coulomb barrier \cite{xu02}.
As a white dwarf collapses to a state with nuclear or supranuclear
densities, strange quark matter seeds may help to trigger the
transition from normal matter to quark matter. However, the seeds
may not be necessary, since the transition could occur
automatically at that high density.

In the core collapse case, the total photon energy could be much
larger than the energy ($\sim 10^{-2}\mathcal{E}_{\rm g}\sim
10^{51}$ erg), with which the out envelope should be expelled,
since the time scale for a proto- strange star with $T\sim
10^{11}$ is usually more than 1 s.
In the AIC case, the bounding energy of a progenitor white dwarf
with mass $\sim M_\odot$ and radius $\sim 10^9$ cm is $E_{\rm
boun}\sim 3\times 10^{50}$ erg, and a minimum mass $\sim 3\times
10^{-3}M_\odot$ of bare strange stars would release an amount of
$E_{\rm boun}$ if each baryon contributes about $50$ MeV after
conversion from hadron matter to strange quark matter. In this
sense, such explosions to produce low-mass strange stars are
powered by the phase-transition energy, rather than the
gravitational energy. Certainly, the low-limit can be smaller if
the progenitor white dwarf is less massive.
It is then not surprising that AIC may produce low-mass strange
stars as long as a strange quark phase conversion\footnote{
The critical condition may depends on various parameters, e.g.,
the chemical composition, the accretion rate, the thermal history,
the stellar rotation, etc.
} %
can occur in the center of a white dwarf with much high
temperature and density.
The mass of the residual bare quark stars may depend on the
details of combustion of nuclear matter into strange quark matter,
especially on the last detonation surface where the
phase-transition can not occur anymore. Certainly, this surface
would be determined by various microphysics (e.g., how much energy
per baryon is released during the phase conversion?).
In case the kick energy is approximately the same, only solar-mass
millisecond pulsars can survive in binaries since low-mass pulsars
may be ejected by the kick.

Recently, it is suggested by \cite{pop04}, that low-mass compact
objects can only form by fragmentation of rapidly rotating
protoneutron stars, and that such objects should have large kick
velocities.
It is worth noting that AIC-produced low mass compact stars might
not have large kick velocities, which may serve as a possible test
of the models of formation mechanisms.
On one hand, if the kick energy $E_{\rm kick}\sim MV^2$ ($V$ the
kick velocity) is dominantly part of the gravitational energy
$\sim GM^2/R$, then one comes to $V\sim \sqrt{M}$. On the other
hand, if $E_{\rm kick}$ is mainly part of phase-transition energy
$\propto M$, then low-mass stars could have similar $V$ as that of
strange stars with $\sim M_\odot$.
Fragmentation might hardly occur for quark stars due to the
color-confinement.

Additionally, AIC-created pulsar-like stars may help to explain
various astrophysical phenomena, the recent work including the
kick velocities of millisecond pulsars \cite{tb96}, the
$r$-process nucleosynthesis of heavy (baryon number $A>130$)
nuclei \cite{fryer99,qw03}, the ultra-high-energy protons
accelerated in the pulsar magnetospheres \cite{pl01}, as well as
the numerical calculations of millisecond pulsar formation in
binary systems \cite{tvs00}.
Nonetheless, besides the high-mass stars, it is interesting to
determine {\em whether an LI-star} (a low- and intermediate-mass
star, with mass $\sim 2\leq M/M_\odot \leq \sim 8$) {\em can die
via a violent event} (e.g., supernova) and maybe produce a
low-mass strange star after unstable nuclear explosion, by detail
calculations and/or observations.

2. {\em The origin of pulsar magnetic fields}.
The different values of $\mu_{\rm m}$ (or $\mu_{\rm v}$) of normal
pulsars and millisecond pulsars (Table 1) may result from their
dissimilar physical processes at birth (iron-core collapse or
AIC).
The magnetic momentum per unit baryon, $\mu_{\rm b}$, could be
dependent on the number, $n$, of quarks of each cluster.
It is suggestive that $\mu_{\rm b}$ of pulsars created by
core-collapse is bigger than that by AIC according to Table 1.
A nucleon has a magnetic momentum of about nuclear magneton,
$\mu_{\rm N}=5\times 10^{-24}$ erg/G, and the corresponding value
$\mu_{\rm m}\sim \mu_{\rm N}/m_{\rm p}\sim 3$ G$\cdot {\rm
cm}^3\cdot$g$^{-1}$, which is much larger than that observed in
pulsars. This hints that the quark clusters in solid quark stars
may have a magnetic momentum per baryon to be $\sim 10^{-(4\sim
6)}$ orders weaker than that of nucleons.

\section{Conclusions and Discussions}

General properties of both rotation- and accretion-powered
low-mass bare strange stars are presented. It is suggested that
normal pulsars with $\sim M_\odot$ masses are produced after
core-collapse supernova explosions, whereas millisecond pulsars
with $\sim (0.1-1) M_\odot$ (and even lower) masses could be the
remains of accretion-induced collapses (AIC) of massive white
dwarfs.
These different channels to form pulsars may result in two types
of ferromagnetic fields: weaker for AIC ($\mu_{\rm m}\sim 10^{-6}$
G$\cdot {\rm cm}^3\cdot$g$^{-1}$) while stronger for core-collapse
($\mu_{\rm m}\sim 10^{-4}$ G$\cdot {\rm cm}^3\cdot$g$^{-1}$).
We note that the low-mass quark stars involved in this paper are
also with small radii, which may be distinguished from low-mass
quark stars with large radii \cite{ahv04}.

Some potential astrophysical appearances relevant to low-mass bare
strange stars are also addressed. We suggest that the radio-quiet
central compact object, 1E 1207.4- 5209, is a low-mass bare
strange star with polar surface magnetic field $\sim 6\times
10^{10}$ G and likely a few kilometers in radius, and it is now at
a critical point of subsonic propeller phase, $P\sim P_{\rm br}$,
in order to understand its timing behavior. A newborn low-mass
strange star could rotate very fast, even with a super-Kepler
frequency. The radius of the dim thermal object, RX J1856.5-3754,
is $R>0.1$ km if its soft UV-optical component radiates from a
spherically quasi-static atmosphere around. It is proposed that
some of the transient unidentified EGRET sources may result from
the collisions of two low-mass strange stars.
It worth noting, in our sense, that the so called Massive Compact
Halo Objects, discovered through gravitational microlensing
\cite{alcock93}, could also be probably low-mass quark stars
formed by evolved stars, rather than quark nuggets born during the
QCD phase transition of the early Universe \cite{ban03}.

The mass of strange quark matter could be as low as of a few
hundreds of baryons (strangelets). Strangelets can be evaporated
through bare surfaces of new-born quark stars, or produced during
collision of two (low-mass) quark stars. Strangelets with $\sim
10^{8-9}$ baryons could be detected as ultra-high energy cosmic
rays \cite{xw03}. In this sense the discrepancy between the
observational fluxes \cite{esa} of AGASA and of HiRes/Fly's Eye
might be explained, since solid strangelets at initially low
temperature should be heated enough to ionize the atmosphere and
would result thus in low radiation of fluorescence in the later
detector.

Can we confirm the small radius of a low-mass bare strange star by
a direct observation of future advanced space telescopes? This
work might be done by the next generation Constellation X-ray
telescope (to be launched in 2009-2010), which covers an energy
band of (0.25-100) keV.
The radii, $R$, of neutron stars are generally greater than 10 km
($R$ of $0.1M_\odot$ mass neutron stars is $\sim 160$ km). If
pulsars are neutron stars, their surfaces should be imaged by the
Constellation-X with much high space resolution, as long as the
separation between the four satellites is greater than $\sim
\lambda d/R\sim 3\lambda_{-8}d_{\rm 100pc}/R_6$ km (Note: the
wavelength of X-ray photon with 10 keV is $\lambda\sim 10^{-8}$
cm, the distance to a neutron star is $d=d_{\rm 100pc} \times
100$pc).
However, if these objects are bare strange stars with low masses,
Constellation-X may not be able to resolve their surfaces.

We suggest also that more electron cyclotron lines could be
detected by future telescopes. The field strength allowing an
absorption detectable in Constellation-X is accordingly from $\sim
2\times 10^{10}$ G to $\sim 9\times 10^{12}$ G, while that for
UVISS (its spectroscopy in two ranges: 125-320 nm and 90-115 nm)
is from $\sim 3\times 10^{10}$ G to $\sim 10^{11}$ G.
The detection of lines in radio pulsars (especially in low-field
millisecond ones) is interesting and important. We may distinguish
neutron or strange stars through constraint of the mass-radius
relations, by knowing the cyclotron-determined $B$ and the timing
result $\sqrt{P{\dot P}}$ [e.g., see Eq.(\ref{b-lowmass}) for
low-mass bare strange stars], in case that the lines form just
above the stellar surfaces.

The properties of radio emission from millisecond pulsars are
remarkably similar to those of normal pulsars \cite{dick92},
although the inferred polar fields range about 4 orders, based on
Eq.(\ref{b}). However, their fields range only $\sim 2$ orders
when the ingredient of mass-changing is included, according to
Table 1 and Eq.(\ref{B-lowmass}), if pulsars are actually strange
stars.
Why has no millisecond pulsar with characteristic age $T_{\rm
c}<10^8$ years been found (or why are most millisecond pulsars so
old)? The answer could be that the initial periods, $P_0$, of
millisecond pulsars spread over a wide range of $\sim 1$ ms (or
even smaller) to $\sim 50$ ms, so that $P$ is not much larger than
$P_0$ [it is thus not reasonable to estimate age by $T_{\rm
c}=P/(2{\dot P})$].
Such a distribution of $P_0$ may be relevant to their birth
processes of AIC.

Can a core-collapse supernova also produce a low-mass bare strange
star? This possibility could not be ruled out in principle. Likely
astrophysical hints could be that the thermal X-ray emission and
rotation power of such a star should be lower than expected
previously. Additionally, the cooling history of a low-mass
strange star should be significantly different from that of
solar-mass ones.
Observationally, two isolated ``low-field'' weak radio pulsars
could be low-mass normal pulsars \cite{lorimer04}: PSR J0609+2130
($P=55.7$ ms, ${\dot P}=3.1\times 10^{-19}$) and PSR J2235+1506
($P=57.9$ ms, ${\dot P}=1.7\times 10^{-19}$).
The polar fields inferred from Eq.(\ref{b-lowmass}) of low-mass
bare strange stars could be higher than that from Eq.(\ref{b}) for
rotation powered pulsars, since one has
\begin{equation}
R_6=2.9\times 10^{15}{\bar B}_{60}B_{12}^{-2}P{\dot P},
\end{equation}
from Eq.(\ref{b-lowmass}) and Eq.(\ref{mr}). We can therefore
obtain radii of rotation-powered pulsars in case that electron
cyclotron absorptions from their surfaces are detected by advanced
X-ray spectrometry.
If these two pulsars have polar fields $B=5\times 10^{10}$ G, the
radii of PSR J0609+2130 and PSR J2235+1506 are thus 2 km and 1 km,
respectively. The very weak radio luminosity might be due to a
very small rotation energy $I\Omega^2/2\propto R^5$ and a small
potential drop $\phi$ in Eq.(\ref{phi}). In this sense, many
low-mass bare strange stars may be not detectable in radio band.

It is sincerely proposed to search low-mass bare strange stars,
especially with masses of $\sim (10^{-1}-10^{-3})M_\odot$, by
re-processing the timing data of radio pulsars.
The systemic long-term variation of timing residual of the
millisecond pulsar (B1937+21) might uncover a companion star with
mass $\sim 10^{-2}M_{\odot}$ \cite{gong03}.
The companion masses of pulsar/white-dwarf binaries are {\em
estimated} to be a few $0.1M_\odot$ \cite{tc99}. Are all these
companions real white dwarfs (or part of them to be just low-mass
strange stars)? Only part of the companions (a few ten-percents)
of pulsar/white-dwarf systems have been optically detected. A
further study on this issue is then surely necessary.

Finally, it is very necessary and essential to probe strange quark
stars through various observations of millisecond-pulsar's
environments (planets, accretion disks), in order to distinguish
these two scenarios on millisecond-pulsar's nature: to be (A)
recycled or (B) supernova-originated.
In case (A), planets and residual accretion disks could be around
such pulsars, but possible mid- or far-infrared emission is still
not been detected \cite{lf04,lww04}. Another point to be not
natural in case (A) is: it is observed that planets orbiting
main-sequent stars can only form around stars with high
metallicities, but the planet captured by PSR B1620-26 is in a
low-metallicity environment (the globular cluster M4). In
addition, the formation of pulsar planets is still a matter of
debate \cite{mh01}.
In case (B), however, observations relevant could be well
understood, since a pulsar (with possibly low mass) and its
planet(s) may born together during a supernova (see the discussion
at the end of \S4 and \cite{xw03}), and no infrared emission can
be detected if no significant supernova-fall-back disk exists.
If the first scenario is right, infrared radiations from both the
disks and the planets\footnote{
Comet-like planets with density of $\sim 1$g/cm$^3$ are much
larger and more grained than that strange planets with $\sim
10^{14}$g/cm$^3$, and contribute consequently more infrared
emission.
} %
could be detectable by the Spitzer Space Telescope and by the
present SCUBA-1 or future -2 detectors of JCMT 15-m ground
telescope. But if the later is true, only negative results can be
concluded.
Surely, these are exciting and interesting subjects to be proposed
when these advanced telescopes operate.
Besides the radio-loud pulsars, it is also valuable to detect
sub-mm emission from radio-quiet pulsar-like compact objects
discovered in high-energy X-ray bands, in order to find hints of
quark stars. In the model presented, compact center objects (CCOs,
e.g., 1E 1207.4-5209) and dim thermal Neutron stars (DTNs, e.g.,
RX J1856.5-3754) might have sub-mm radiation from the cold
material around the stars; but no sub-mm emission is possible if
no accretion occurs there. It is then very interesting to test and
constrain the models through observations of CCOs and DTNs in
sub-mm wavelengths.

In conclusion, if pulsar-like stars are strange quark stars, part
of them should consequently be of low-masses unless one can
convince us that no astrophysical process results in the formation
of low-mass quark stars.
Since they are also X-ray emitters, we may expect that some of
them could have been detected by {\em Chandra} or {\em XMM-Newton}
(e.g., the Chandra $\sim 1$ Ms X-ray survey).


\section*{Acknowledgments}

I would like to acknowledge the discussions of Dr. X. D. Li (for
the propeller phase) and Dr. Bing Zhang (for pulsar death lines),
and to thank various stimulating discussions in the pulsar group
of Peking university.
This work is supported by National Nature Sciences Foundation of
China (10273001) and the Special Funds for Major State Basic
Research Projects of China (G2000077602).

\label{lastpage}

\end{document}